\documentstyle[aps,multicol,epsfig]{revtex}

\title{Multi-particle entanglement of hot trapped ions}

\author{Klaus M\o lmer \cite{klaus} and Anders S\o rensen \cite{anders}\\
 {\small Institute of Physics and Astronomy, University of Aarhus,\\
DK-8000 \AA rhus C, Denmark}}
\begin{document}
\draft

\maketitle

\begin{abstract}
We propose an efficient method to produce  multi-particle
entangled states of ions in an ion trap
for which a wide range of interesting effects and applications 
have been suggested. Our preparation scheme exploits
the collective vibrational motion of the ions,
but it works in such
a way that this motion need not be fully controlled in the
experiment. The ions may, {\it e.g.}, be in thermal motion and 
exchange mechanical energy with a surrounding heat bath without
detrimental effects on the internal state preparation. Our scheme does
not require access to the individual ions in the trap.
\end{abstract}


\widetext
\tighten
\begin{multicols}{2}

We describe in this Letter a method to construct states of the form
\begin{eqnarray}
|\Psi\rangle = {1\over \sqrt{2}}[{\rm e}^{i\phi_g}|gg ... g\rangle + 
{\rm e}^{i\phi_e}|ee ... e\rangle],
\label{GHZ}
\end{eqnarray}
where $|gg...e\rangle$ and $|ee...e\rangle$ ($N$
terms) are product states describing $N$ ions
which are all in the (same)
internal state $g$ or $e$.
The unique  correlation between different particles in such an 
entangled state is a significant non-classical feature of quantum 
mechanics.
Entangled states with two particles  have been
employed to test Bell's inequality and to 
rule out local realistic descriptions of Nature \cite{Aspect}. 
States like (\ref{GHZ}) with three particles, the so-called GHZ states
\cite{GHZ}, and with more particles allow a closer scrutiny of the 
role of quantum correlations, and there is a large current interest 
in carrying out experiments on such states.
It has also been shown that the states
(\ref{GHZ}) are ideal for spectroscopic investigations
\cite{Spin}, and the study  of entangled states of many
particles is in line with the rapidly growing interest in quantum
encoding and quantum information processing \cite{QC}.
Together with an analysis of their  applicability within high precision
frequency measurements,
a method to produce such states in an ion trap was proposed in 
\cite{Spin}. 
That method requires experimental control of two collective 
vibrational modes of the ions,
which are involved in stepwise  manipulations of the
internal states of the ions. 

In order to describe our more straightforward proposal,
we shall show that the desired state (\ref{GHZ}) can be  obtained
by application of one simple interaction Hamiltonian.
We shall then show that  this Hamiltonian is realized in an ion trap
by simply illuminating all the ions with laser beams with two different 
colours, and we shall present numerical results that display the performance
of our proposal under quite relaxed assumptions for the functioning of
the ion trap.

We begin with all ions in the ground state $|g\rangle$, and 
we assume that our external perturbation interacts in the same way
with all ions, so that the joint internal state 
remains symmetric with respect to interchange 
of the ions.
In such a situation, it is convenient to apply the 
spin representation of atomic states, where a collective
state of all ions is represented by
$|J,M\rangle$ which is an eigenstate of the $J_z$ operator,
$J_z|J,M\rangle = \hbar M|J,M\rangle$, $M=-J, ... ,J$, where
$\hbar$ is Planck's constant.
$J_z$ "counts" the number of excited ions, so that
a state, invariant under permutation of the ions,
with a certain number $N_e$ of excited ions is described as
$|J=N/2,M=N_e-N/2\rangle$, and 
the state (\ref{GHZ}) is a superposition 
of the states $|N/2,-N/2\rangle$ and $|N/2,N/2\rangle$. 
An increase of the quantum number $M$ by unity corresponds to a single
ionic excitation, and it
is described by the angular momentum ladder operator
$J_+|J,M\rangle = \hbar \sqrt{(J+M+1)(J-M)}|J,M+1\rangle$,
where the normalization factor accounts precisely for
the combinatorial aspects of this excitation of the system.
As we shall detail below, it is possible to apply classical
laser irradiation which 
cannot be absorbed by a single ion, but which requires two ions to
undergo the $|g\rangle \leftrightarrow |e\rangle$ transition,
simultaneously. 
When this field is applied to all ions in the trap,
the effective Hamiltonian describing the
evolution of the state of the ions in the spin representation
can be written,
\begin{eqnarray}
H={\chi \over \hbar}(J_+^2+J_-^2+J_+J_-+J_-J_+) = 4 {\chi \over \hbar} J_x^2
\label{Ham}
\end{eqnarray}
where $J_-$ is the hermitian conjugate of $J_+$ and 
where $J_x=(J_++J_-)/2$ denotes the 
$x$-component of the effective spin.

Time evolution is accommodated by application of the operator
$U(t) = \exp(-i 4\chi J_x^2 t/\hbar^2)$ to
the initial state $|N/2,-N/2\rangle$ and the population becomes distributed
on all $|J,M\rangle$ states, with $M$ differing from $-N/2$ by an even
number. If $N$ is even, population is transferred 
all the way to the state $|N/2,N/2\rangle$, which is the second
component of (\ref{GHZ}). Moreover, at the
instant $t=\pi/(8\chi)$, the state is precisely of the form (\ref{GHZ}),
with $\phi_g=-\pi/4$ and $\phi_e=\pi/4+N\pi/2$. 
We show in Fig. 1. the time evolution of the populations of the
two extremal states for different (even) numbers of ions. 
At $t=\pi/(8\chi)$ and at later odd
multiples of this value, both populations
equal one half. 

This result can be understood from the rotation properties of 
angular momenta:
The initial state can be expanded on eigenstates of $J_x$:
$|N/2,-N/2\rangle=\sum_M c_M |N/2,M\rangle_x$, and
from the properties of the Wigner rotation functions \cite{Brink},
it follows that $|N/2,N/2\rangle=\sum_M c_M (-1)^{N/2-M} |N/2,M\rangle_x$.
Inserting the values for $\phi_g$ and
$\phi_e$ mentioned above, we can therefore write the state (\ref{GHZ})
as $\sum_M c_M {1\over \sqrt{2}}({\rm e}^{-i\pi/4}+(-1)^M{\rm
e}^{i\pi/4})|N/2,M\rangle_x$. The net factors multiplying 
the initial amplitudes $c_M$ are
unity for $M$ even and $-i$ for $M$ odd.
The action of $U(t)$ in the $J_x$ basis amounts to a multiplication of
each amplitude $c_M$ by $\exp(-4i\chi M^2 t)$, and for $t=\pi/(8\chi)$
this factor just attains the desired value of unity for 
$M$ even and $-i$ for $M$ odd.

Note that the state (\ref{GHZ}) is not only a multi-particle entangled
state; it is a superposition of two ``mesoscopically distinguishable''
states. In Quantum Optics, such states have
been studied experimentally for single quantum systems,
such as a harmonically trapped ion\cite{WinCat} and a single mode of the 
quantized radiation field \cite{Haroche}, 
excited into superpositions of states of ``mesoscopic'' separation.
Mathematically, our analysis above  resembles the one for production of
such superpositions by  propagation of an optical field through an
amplitude dispersive medium \cite{Yurke}.

We now turn to a ``microscopic" analysis of 
a collection of $N$ two-level ions in a linear ion trap. 
The choice of a linear trap eliminates the so-called micro-motion
for ions on the axis, and by choosing the transverse
confinement much stronger than the longitudinal one, we
can neglect coupling to transverse vibrations.    Due to the
Coulomb repulsion among the ions, the various longitudinal vibrational
modes  have different frequencies and by choosing the laser detuning close
to the center-of-mass vibrational
frequency we may exclude other modes from our analysis \cite{othermode}.  

The quantized energy of the total system is composed of the 
center-of-mass vibrational energy of the ion string and  the
internal electronic energy of the ions.
The vibrational energy is 
represented by a harmonic oscillator with ladder operators 
$a$ and $a^\dagger$ and frequency
$\nu$, and the internal energy is
described by Pauli matrices $\sigma_{zi}$ and the
energy difference $\hbar \omega_{eg}$, 
\begin{eqnarray}
 H_0&=&\hbar \nu (a^{\dagger} a+1/2)+
{\hbar \omega_{eg}\over 2}\sum_i \sigma_{zi}.
\label{ham0}
\end{eqnarray}
The ions interact with two laser fields with
frequencies $\omega_j$ ($j=1,2$). We assume that both fields have the same
intensity and we also assume that the ion-laser interaction strength
characterized by the Rabi frequency $\Omega$  is the same for all ions. 
Pauli matrices
$\sigma_{\pm i}$ describe internal atomic transitions
and the interaction Hamiltonian can be written
\begin{eqnarray}
 H_{int}&=&\sum_{i,j} \frac{\hbar\Omega}{2}
 (\sigma_{+i}e^{i(\eta_j(a+a^{\dagger})-\omega_j t)}+ h.c.).
 \label{hamilton}
\end{eqnarray}
($h.c.$ in (\ref{hamilton}) denotes the hermitian conjugate.)
The Dicke parameters $\eta_j$ are products of the projection of
the $k$-vector of the $j$'th laser field along the direction of the
string of ions and the {\it rms} excursion of the 
ionic center-of-mass along this direction \cite{raman}. In practical
realizations the $k$-vectors can be nearly identical for the two fields,
and we assume 
$\eta_1=\eta_2=\eta$.
Numerical solutions of the
Schr\"odinger equation with the Hamiltonian given by
Eqs.(\ref{ham0},\ref{hamilton}), are readily obtained
within the basis of 
$2^N$ internal states and  a sufficiently large set of 
vibrational states $\{|f_1f_2 ... f_N\rangle|n\rangle\}$, $f_i=e$ or $g$.
First, however, we discuss a perturbative analysis pointing to appropriate
choices of parameters which will effectively yield the spin Hamiltonian
(\ref{Ham}). 

The energy
levels of the ion string constitute a ladder of equidistant
vibrational levels added to the internal ionic energies.
This ladder structure has been extensively used in laser 
manipulation of ions in
traps, where laser frequencies 
differing  from the atomic transition
frequency by a multiple of the trap frequency may be used to
control the internal and external dynamics \cite{WinCat}, and excitation of
collective  vibrational modes for several ions has been
proposed as a means for implementing quantum logic in an ion trap quantum
computer \cite{cirac}. 

Here, we make use of the idea we put forward in \cite{prl} to apply laser
fields with two  different frequencies 
so that the two-photon process, exciting any pair of ions in the
trap
$|gg\rangle \rightarrow |ee\rangle$ is resonant, {\it i.e.},
$\omega_1+\omega_2=2\omega_{eg}$, but neither of the frequencies
are resonant with single excitations of the ions.
By choosing $\omega_1=\omega_{eg}-\delta$ and
$\omega_2=\omega_{eg}+\delta$, where $\delta$ is close to, but not 
resonant with, the center-of-mass vibration frequency we may limit our  perturbative
analysis to involve only the  intermediate
states with one excited ion and a 
vibrational quantum number raised or lowered by unity.
Note that the lack of degeneracy of vibrational modes is crucial because it
prevents a number of 
modes to interfere and obstruct the collective transitions in the
multi-particle system.
We show in Fig. 2 the
effective transition paths between states 
$|gg\rangle|n\rangle$ and $|ee\rangle|n\rangle$ 
and between $|eg\rangle|n\rangle$ and $|ge\rangle|n\rangle$ mediated by the
interaction with the laser fields within 
any pair of ions.
Although they serve as intermediate states in the process, the
intermediate states are only virtually {\it i.e.}, negligibly, populated. 
The interaction strengths of the
different one-photon transitions in Fig. 2 depend on the value of the
vibrational quantum number $n$ and in the limit
of $\eta\sqrt{n+1}<<1$, terms between states $n$ and 
$n+1$ are proportional to $\sqrt{n+1}$ whereas terms between $n-1$ and $n$
are proportional to $\sqrt{n}$. When we add the  contributions
from the four different transition paths shown in either panel of Fig. 2,
a ``miracle" occurs: The paths 
(virtually) involving intermediate states with $n+1$ vibration quanta 
yield $(\sqrt{n+1}\Omega\eta)^2/(\delta-\nu)$ and the
ones with $n-1$ quanta yield
$(\sqrt{n}\Omega\eta)^2/(\nu-\delta)$. The signs of the denominators cause a destructive
interference which exactly removes the $n$-dependence from the
total effective Rabi frequency of the two-photon transition. This implies 
that the internal state evolves in a manner independent of the
external state, which may be in a mixed ({\it e.g.},
thermal) state and  which  may even evolve with time.

When the summation over all ions is carried out, the couplings
lead to the spin Hamiltonian (\ref{Ham}) with
\begin{equation}
  \chi=\frac{\eta^2\Omega^2\nu}{2(\nu^2-\delta^2)},
\end{equation}
and to the 
extent that our perturbative treatment is valid, we therefore
expect to realize the results, presented in Fig. 1. 
In Fig. 3. we show numerical simulations of the dynamics
governed by the correct Hamiltonian (\ref{ham0},\ref{hamilton}) for 4 ions, 
{\it i.e.} beyond perturbation theory and 
to all orders in the Dicke parameter $\eta$. We
assume a distribution on vibrational levels with
quantum numbers ranging between 0 and 40, and we
assume that the center-of-mass motion exchanges 
energy with a surrounding thermal reservoir represented by relaxation
operators $c_1=\sqrt{\Gamma(n_{th}+1)}a$ and $c_2=\sqrt{\Gamma
  n_{th}}a^\dagger$.  Our calculation for a reservoir with a mean 
excitation  of $n_{th}=5$ vibrational quanta at the trap frequency 
and a coupling strength $\Gamma=0.0001$ may represent a heating mechanism
towards higher temperatures  
counteracted by laser cooling on one ion in the
string, reserved for this  purpose. The action of the reservoir 
is described by quantum jumps and a non-Hermitian Hamiltonian  in a Monte
Carlo  wave function
simulation \cite{MCWF}. Part (a) of the figure
represents the result of a single Monte Carlo realisation with 
quantum jumps in the vibrational states at instants of time, indicated
by arrows in the figure. Part (b) is an average over 10 such 
realisations. The full lines present the populations and
coherences of the internal state of the ions, obtained by  tracing out
the vibrational degrees of freedom. For comparison we also show the result
of a calculation with the corresponding spin Hamiltonian (\ref{Ham})
(dashed curve).   With the realistic parameters chosen, the state at $\nu t
\sim 1500$ is 
very close to the desired state in (\ref{GHZ}). Small oscillatory deviations 
from the smooth results of Fig. 1. are seen. The magnitude of the
oscillations is given by the physical parameters, so that in the
limit of small $\eta$ and $\Omega$, they can be made arbitrarily small
\cite{prl}.

The pairwise interaction with ions does not produce a coherent
coupling of the two components in (\ref{GHZ}) if $N$ is odd.
In this case, however, one can verify within the spin model,
that it suffices to apply a linear coupling  $H_1=4\xi J_x$ for the
duration $t=\pi/(8\xi)$  
 in addition to application of our quadratic term (\ref{Ham}).
Since $J_x$ and $J_x^2$ commute, the linear Hamiltonian
may actually be applied before, after or simultaneously 
with the quadratic one. This additional  operation is easy
to implement in the ion trap, where one only needs to 
drive the ions on the atomic resonance frequency $\omega_{eg}$.

In summary, we have presented a method to prepare 
a multi-particle entangled state of trapped ions.
Our suggestion aims at 
experimental application, and it is relevant both 
for experiments with limited control of the vibrational motion, where the
ions heat up for different 
technical reasons, or where there is not  access to individual ions.  The
possibility to exercise control 
over some quantum degrees of freedom of a system while
others are uncontrolled is common place in
coherent optical excitation experiments, where the internal states of atoms
in a gas may be controlled regardless of the external thermal motion. Our analysis
adds the surprising possibility of coherently controlling
the mutual internal state of several particles by explicit
use of  the uncontrolled incoherently excited  degrees of freedom.
We believe that much wider possibilities for quantum state
control become feasible with the introduction of the crucial
elements of our method: resonance conditions allowing only
pairwise transitions,
virtual excitations of un-controlled
degrees of freedom, and partially destructive interference between transition
paths. We have recently applied these ideas to quantum computation in an
ion trap
\cite{prl} and we
are currently investigating their possible use in other physical systems.

\begin{figure}
\begin{center}
 \epsfig{height=3.4in,angle=270,file=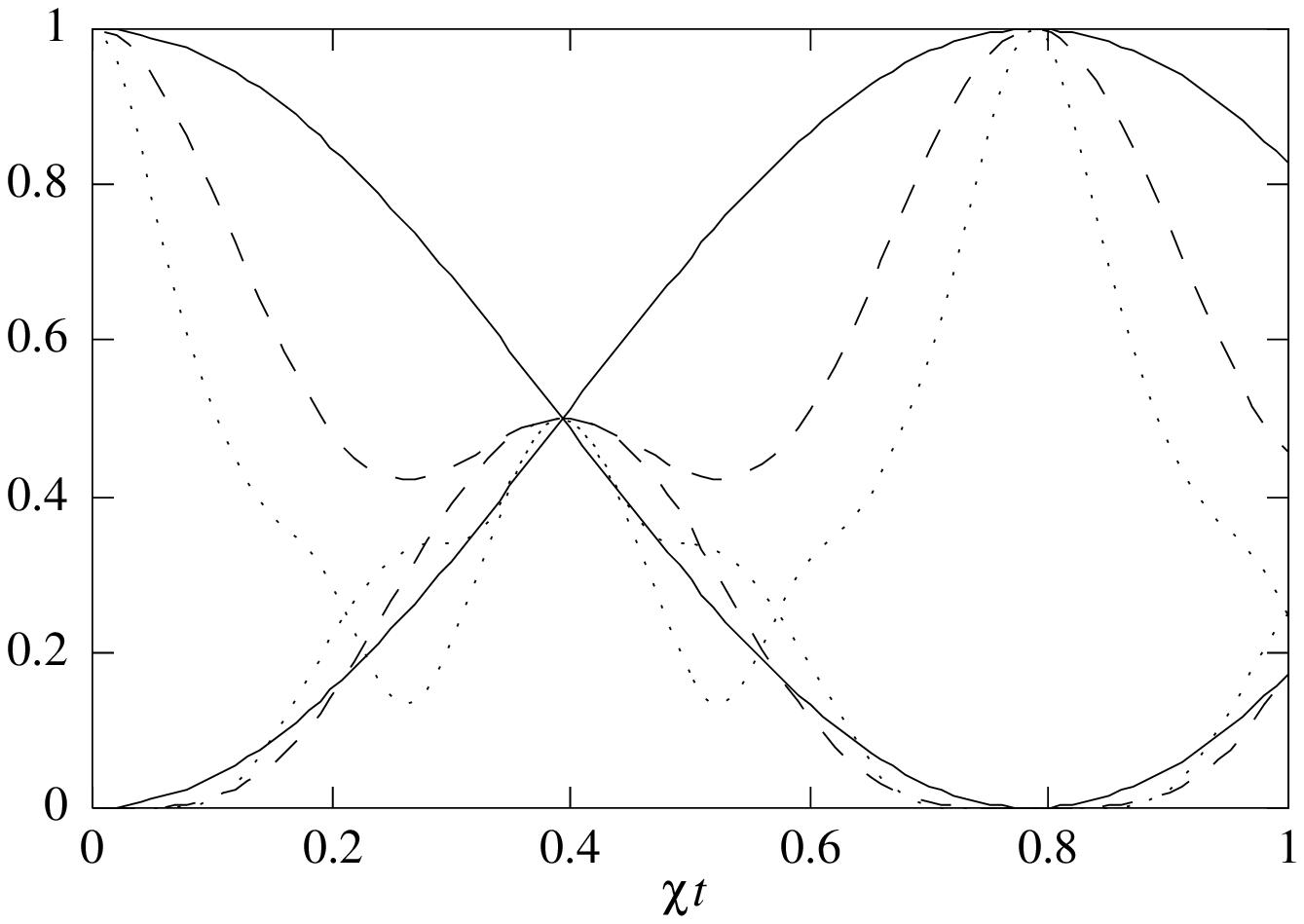}
\end{center}

\caption{Time evolution of the 
population of the joint ionic ground state
$|gg ... g\rangle = |N/2,-N/2\rangle$ (curves starting from the value
of unity at $t=0$), and the joint ionic excited state
$|ee ... e\rangle = |N/2,N/2\rangle$. Results are presented for
different values of the number of ions: N=2 (solid curves),
N=4 (dashed curves) and N=8 (dotted curves).
At $t=\pi/(8\chi)$ the states are in a
50-50 superposition and the state (\ref{GHZ}) is obtained.}
\label{spin}
\end{figure}
\bigskip

\noindent 
\begin{figure}
\begin{center}
 \epsfig{file=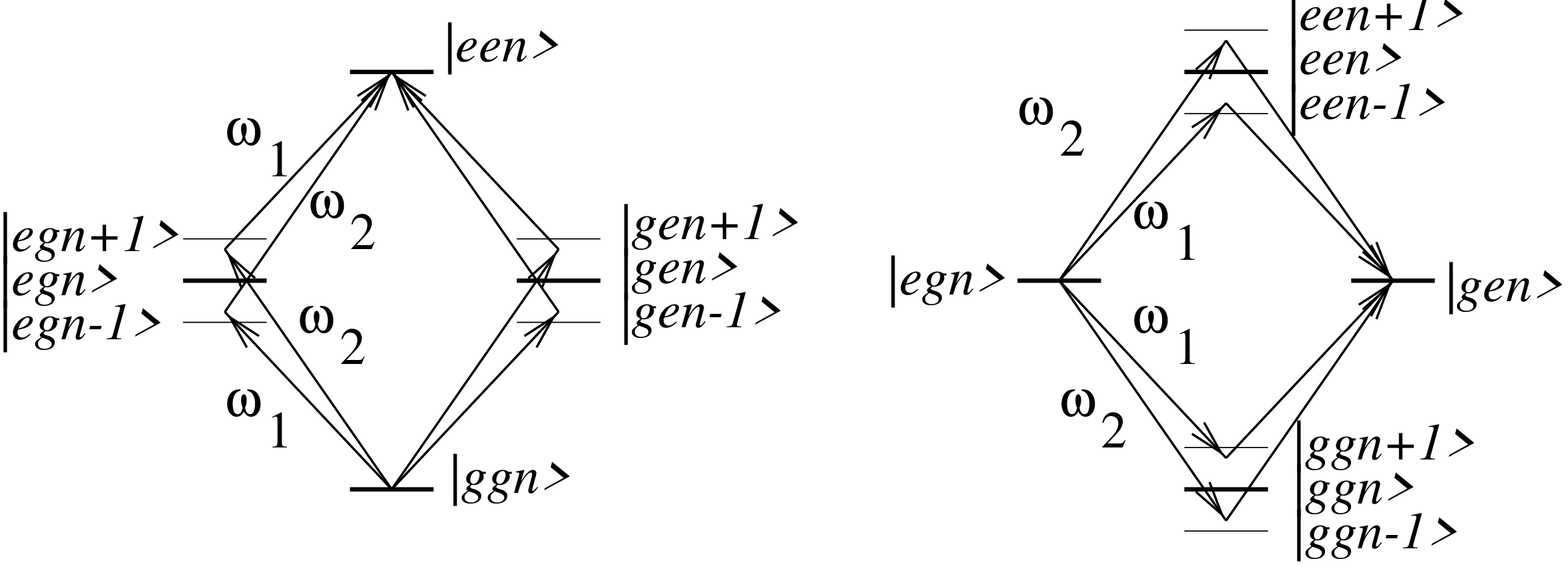,angle=270,width=3.4in}
\end{center}
\caption{Level scheme for a pair of ions sharing an oscillator degree
of freedom. Left: by application of laser light with frequencies
$\omega_{eg} \pm \delta$, where $\delta$ is somewhat smaller than the vibrational
frequency $\nu$, we  identify four transition paths between the
states $|gg\rangle|n\rangle$ and $|ee\rangle|n\rangle$, which interfere
as described in the text. Right: four similar transition paths are
identified between states $|eg\rangle|n\rangle$ and
$|ge\rangle|n\rangle$, yielding the same effective coupling among
these states as between the states in the left panel.}
\label{energy}
\end{figure}

\bigskip
\noindent
\begin{figure}
  \begin{center}
    \epsfig{height=3.4in,angle=270,file=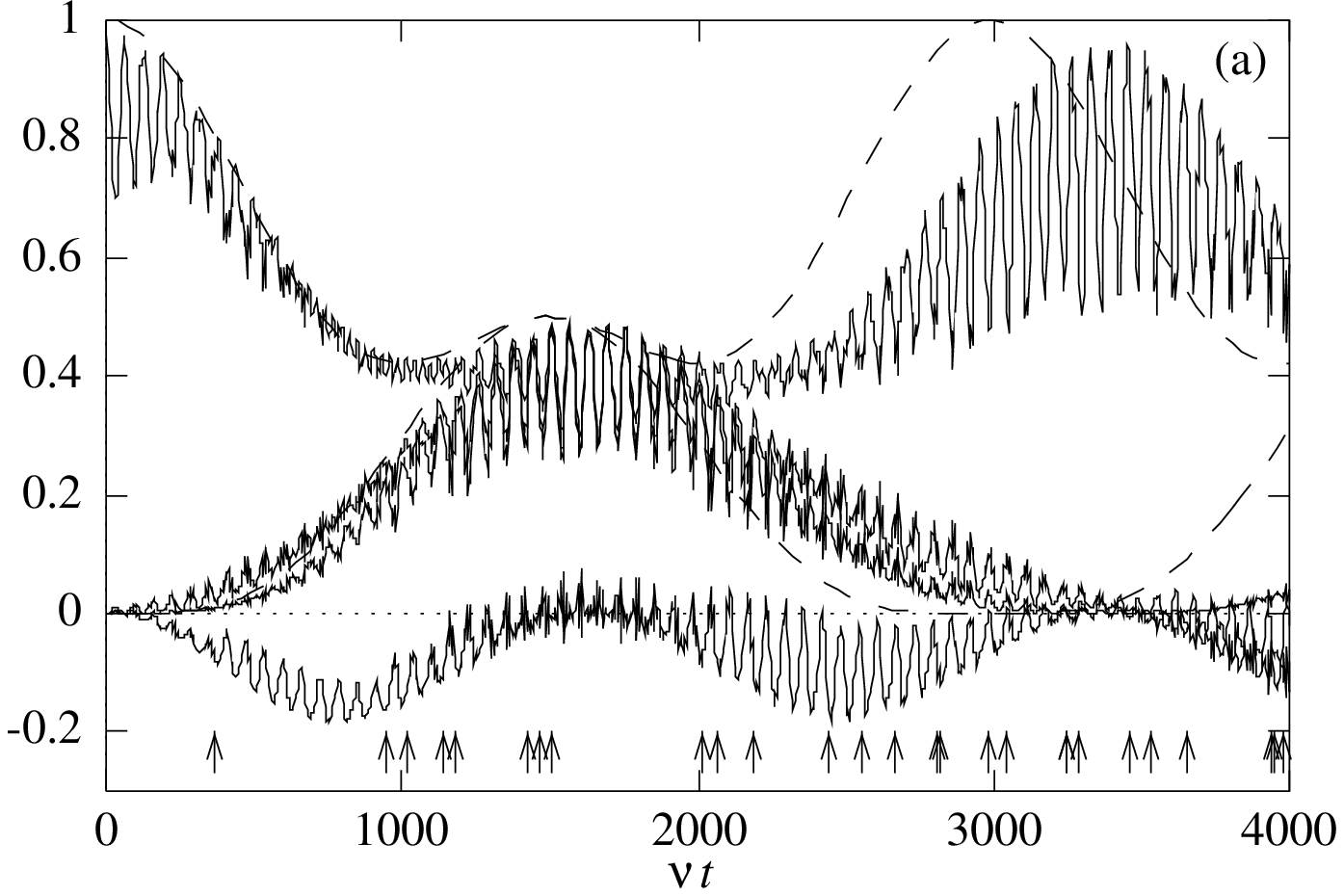}

    \epsfig{height=3.4in,angle=270,file=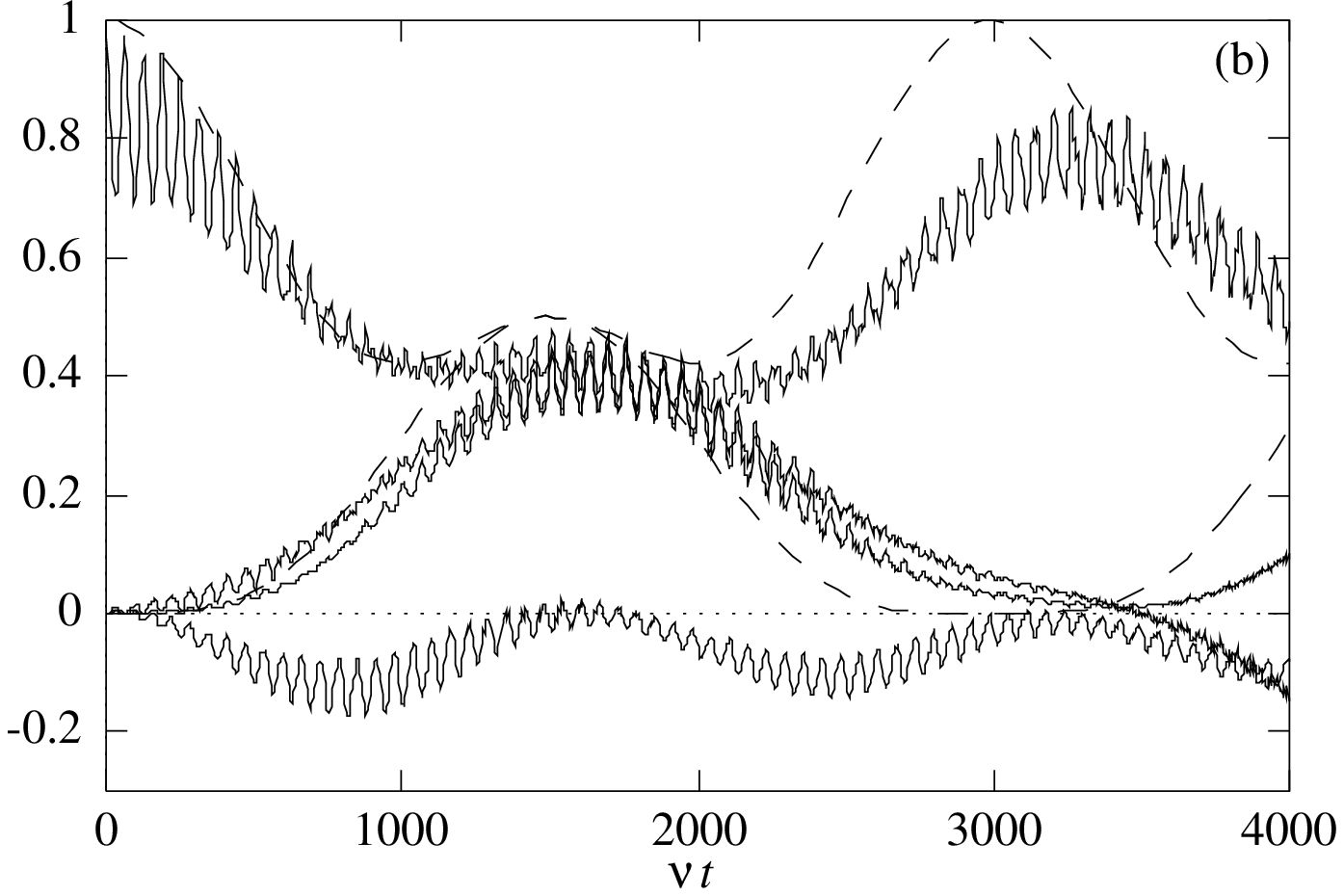}

\end{center}
\caption{Results of numerical integration of the Schr\"odinger equation for 4 ions. The
physical 
parameters are $\delta=0.9 \nu$, $\Omega=0.1\nu$, $\eta=0.1$,
$\Gamma=0.0001\nu$ and $n_{th}=5$. (a) Single Monte 
Carlo realisation with 
quantum jumps in the vibrational states  at instants of time, indicated
by arrows in the figure. (b) Average over 10
realisations.  The first full curve (counted from above at $\nu t<1000$) is
the population  of the joint ground state, the second one 
is the imaginary part of the internal density
matrix element between the joint ground and the joint excited state, the third one is the
population of the joint excited state, and the 
fourth one is the real part of the off-diagonal density
matrix element. The dashed curves are equivalent to the $N=4$ curves of Fig. \ref{spin},
obtained by application of the spin Hamiltonian (\ref{Ham}) with
$\chi=\frac{\eta^2\Omega^2\nu}{2(\nu^2-\delta^2)}$.} 
\label{numerical}
\end{figure}
\end{multicols}

\begin{thebibliography}{99}
\bibitem{othermode} Numerical tests including
an additional vibrational mode prove that
it is sufficient to deal with the near-resonant
center-of-mass mode of the ions.
\end{thebibliography}
\end{document}